# Capacitor-within-Capacitor

H. Grebel
Electronic Imaging Center, NJIT, University Heights, Newark, NJ 07102. grebel@njit.edu

**Abstract:** Capacitors are typically connected together in one of two configurations: either in series, or in parallel. Here we study a capacitor-within-capacitor (CWC). Simulations and experiments indicate that the overall capacitance of the structured cell may be made larger than an ordinary two-plate counterpart by at least 50%. Simulations also indicate that the cell's capacitance may be controlled by the gate voltage. Overall, this concept is deemed suitable for microfluidic sensing and possibly for energy storage elements.

## I. Introduction:

In equivalent circuit terms, capacitors may be connected in series (the overall capacitance becomes smaller than either capacitors), or in parallel (the overall capacitance is the sum of individual capacitors) [1]. Here we consider a third possibility, a capacitor-within-capacitor (CWC). Specifically, one capacitor, the gate capacitor is placed either within (inner gate construction) or outside (outer gate construction) a cell capacitor. The effect is purely linear and relies on induction. Furthermore, such concept is general and may be applied to dielectric (body of the manuscript) and super-capacitors (SI section). Simulations on the control of such structure by the gate voltage are provided in the SI section, as well. One may argue that if a cell capacitor is maintained at a constant voltage, then more energy may be stored by this design.

Simulations of the structure under a short and open inner gate are provided in Section II. Experiments in Section III reveal that one can affect the cell's capacitance by simply connecting the gate electrodes with a variable resistor. Equivalent circuits are provided in Section IV – Discussion. It is worth noting that equivalent circuits do not take the thickness of the electrodes into account; thin gate electrodes do not fully mask the polarization effect of the cell. The effect is scalable; by replacing the dielectric in the capacitor with an ionic liquid, a much larger effect is obtained. Both simulations and experiments demonstrate that the overall cell's capacitance increases upon introducing a gate structure, either outside or inside the cell. This is attributed to an increased cell polarity, leading to a re-distribution of electrical charges on the cell's electrodes. It is, therefore, suggested that one can amass more charges on a low voltage capacitors, realizing a better energy storage.

## II. Simulations:

**Dielectric capacitor – inner gate electrodes:** The dielectric capacitor is shown in Fig. 1. The outer capacitor ($C_{cell}$) is polarized by a potential difference $V_0$ between the cell's electrodes. A finite-element code (Comsol) was used for the simulations. The cell's capacitance is simply the simulated terminal charge divided by the cell's voltage ($V_0$=1 V in our case). As discussed in Section IV.a. Energy Considerations, limiting our attention to the cell's capacitance does not violate any energy conservation law since the energy used for charge separation at the gate is reflected through a change in charge at the cell's electrodes. The structure was assumed to be immersed in a liquid dielectric with a relative dielectric constant of $\varepsilon_r$=100.

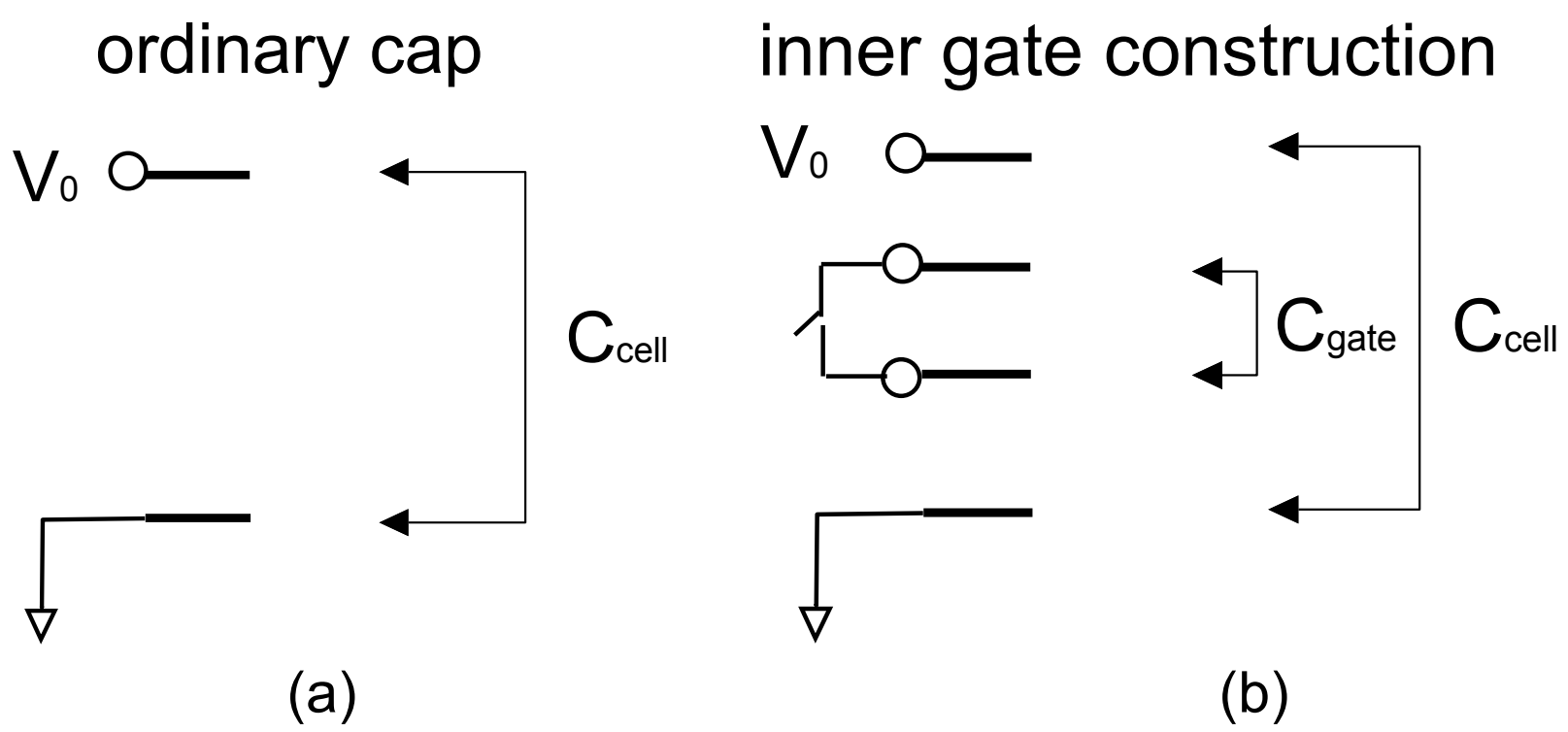

Fig. 1. (a) An ordinary dielectric capacitor. (b) A structure made of a capacitor-within-capacitor.

Let us consider two cases where the gate electrode are floating: in case 1 (Fig. 2a) the gate is open whereas in case 2 (Fig. 2b) the gate is short. Short gate was mimicked with floating edges connecting the two floating gate electrodes. As can be seen from Fig. 2b, the potential distribution with the gate region is constant. The cell's capacitance of the open gate was assessed as 1.44 nF whereas the cell's capacitance with a short gate was 1.86 nF; an increase of 29%.

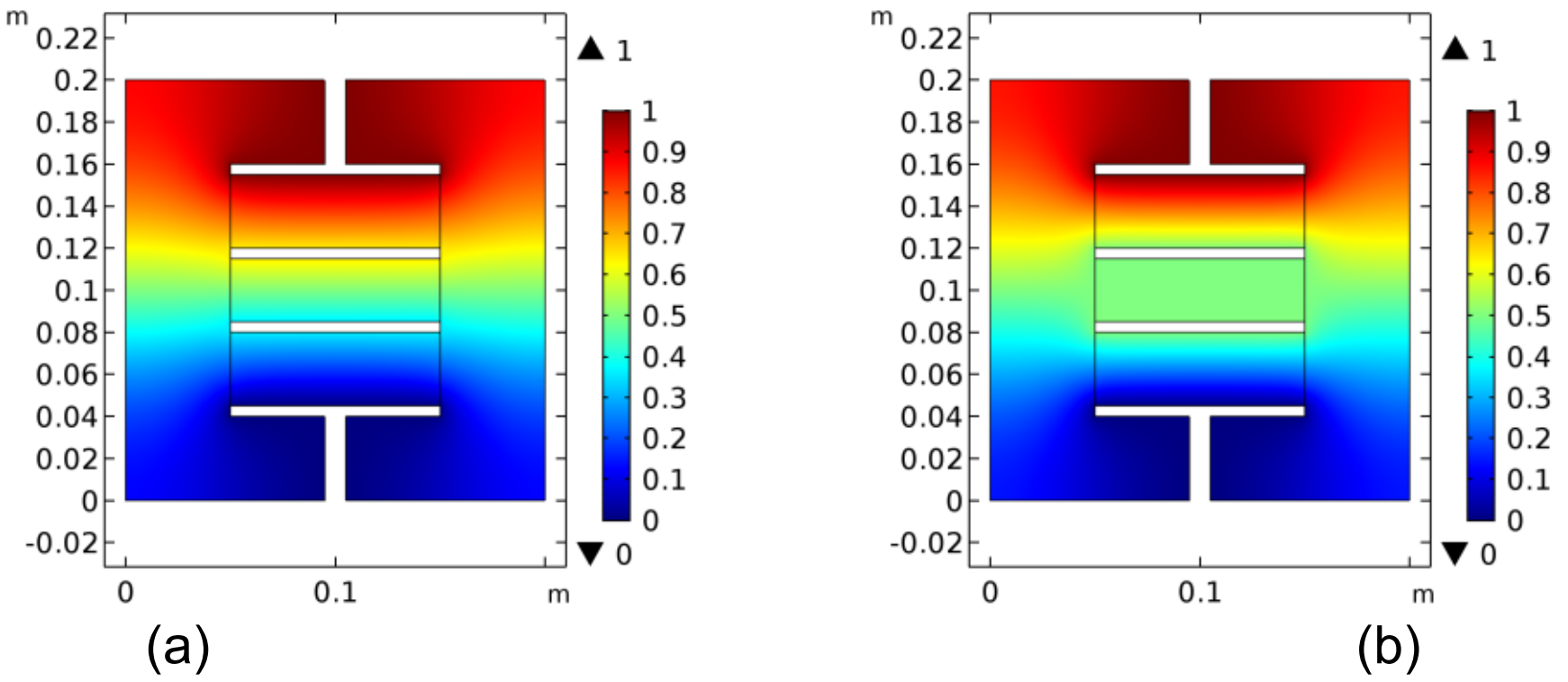

Fig. 2. (a) Potential distribution for a floating and open gate (the two inner electrodes). (b) Potential distribution when the inner gate were short. Short was made by connecting the edges of the gate with a floating line.

## III. Experiments:

Methodology: the 4x4 $cm^2$ electrodes were made out of 25 micron copper foil, with a protruding notch in each plate, left for contact. This way, the thickness of the electrode has not been altered by a wire contact. The dielectric material was a 110 micron thick plastic transparency film and the entire structure was held between two glass plates by two paper clips. A 90 micron soaked lens tissue with ionic liquid (1-n-Butyl-3-methylimidazolium hexafluorophosphate, Thermo-Fisher) served as an electrolyte as well. A hand held multi-meter (BK Precision) and LCR bridge (QuadTech) were used for the capacitance measurements. The simulations were DC calculations whereas a typical capacitance measurement employs a small AC signal. Under small AC signal approximation any independent DC voltage source is treated as a short due to its small internal impedance. Indeed, the measured cell capacitance, with a gate battery in place, was exactly the same as for the case where the two gate plates were short. In order to observe the effect of the gate bias on the cell capacitance at AC frequencies, one needs to synchronize it with the cell's measurements or dope the dielectric layers.

For the AC measurements, an external trigger (not shown) drove the scope and Ground was provided by the function-generator. Since the system may have its own capacitance, a reference trace of only the signal is provided below (Fig. 7). The system capacitance was smaller than the cell's capacitance when including the open gate structure.

***III.a. Cell capacitance with inner gate construction:***
Inner gate configuration is depicted in Fig. 3a. Results obtained with a hand-held unit are shown in Fig. 4b-c.

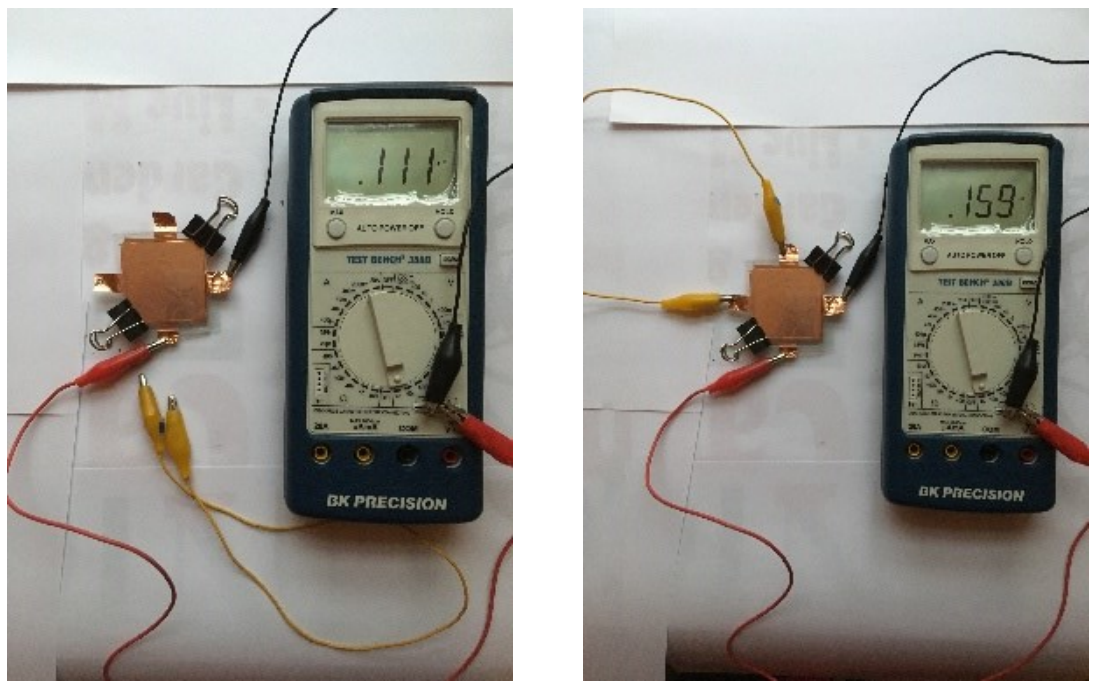

(a) (b)

Fig. 3. (a) Experiment: the inner gate electrodes were open the capacitance was 0.111 nF. (b) Experiment: the inner gate electrodes were short (connected by a yellow jumper). The capacitance was 0.159 nF. Red wire: positive cell's contact; black wire: negative cell's contact; the yellow jumper provides contact between the inner (gate) electrodes. The reading without the gate was similar to those of an open gate structure.

***III.b. Cell capacitance with outer gate construction:***
A complementary design and its related experimental data are provided in Figs 4.

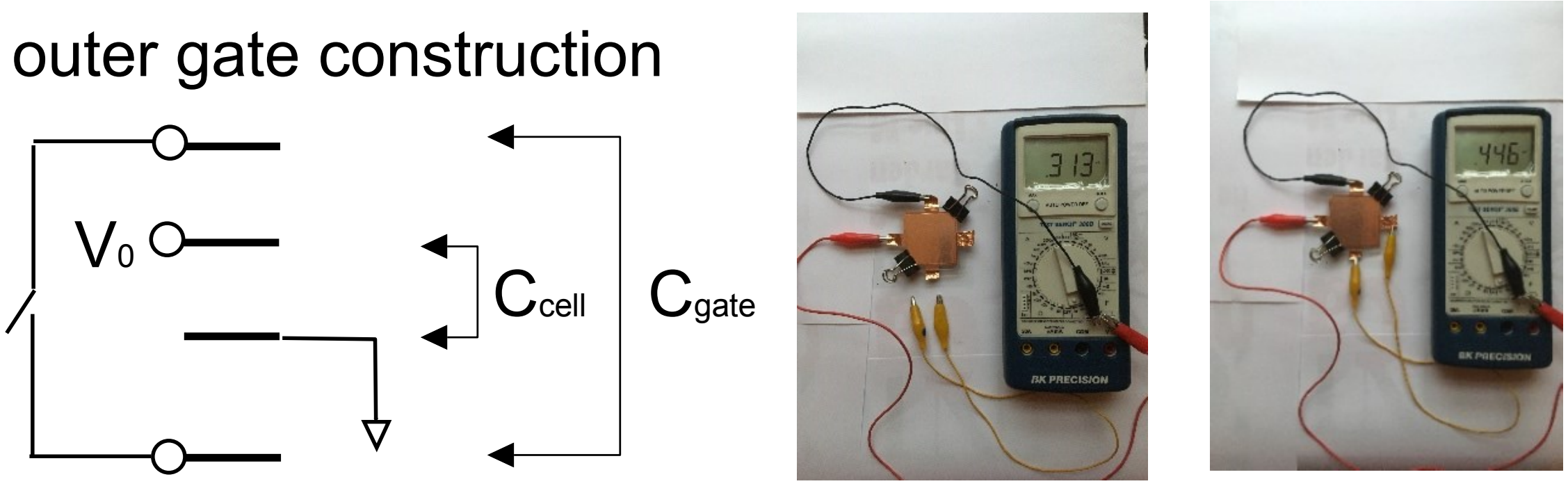

(a) (b) (c)

Fig. 4. (a) Reversing the roles of capacitor-within-capacitor: the cell capacitor is defined by the inner electrodes set whereas the gate capacitor is constructed by the outer electrode set. (b) Experiment: the outer gate electrodes were open. The capacitance was 0.313 nF. (c) Experiment: the outer gate electrodes were short (connected by a yellow jumper). The capacitance was 0.446 nF. Red wire: positive cell's contact; black wire: negative cell's contact; the yellow jumper provides contact between the outer (gate) electrodes.

Capacitance as a function of the gate's shunt resistance is shown in Fig. 5. As the resistance increases, the cell's capacitance decreases.

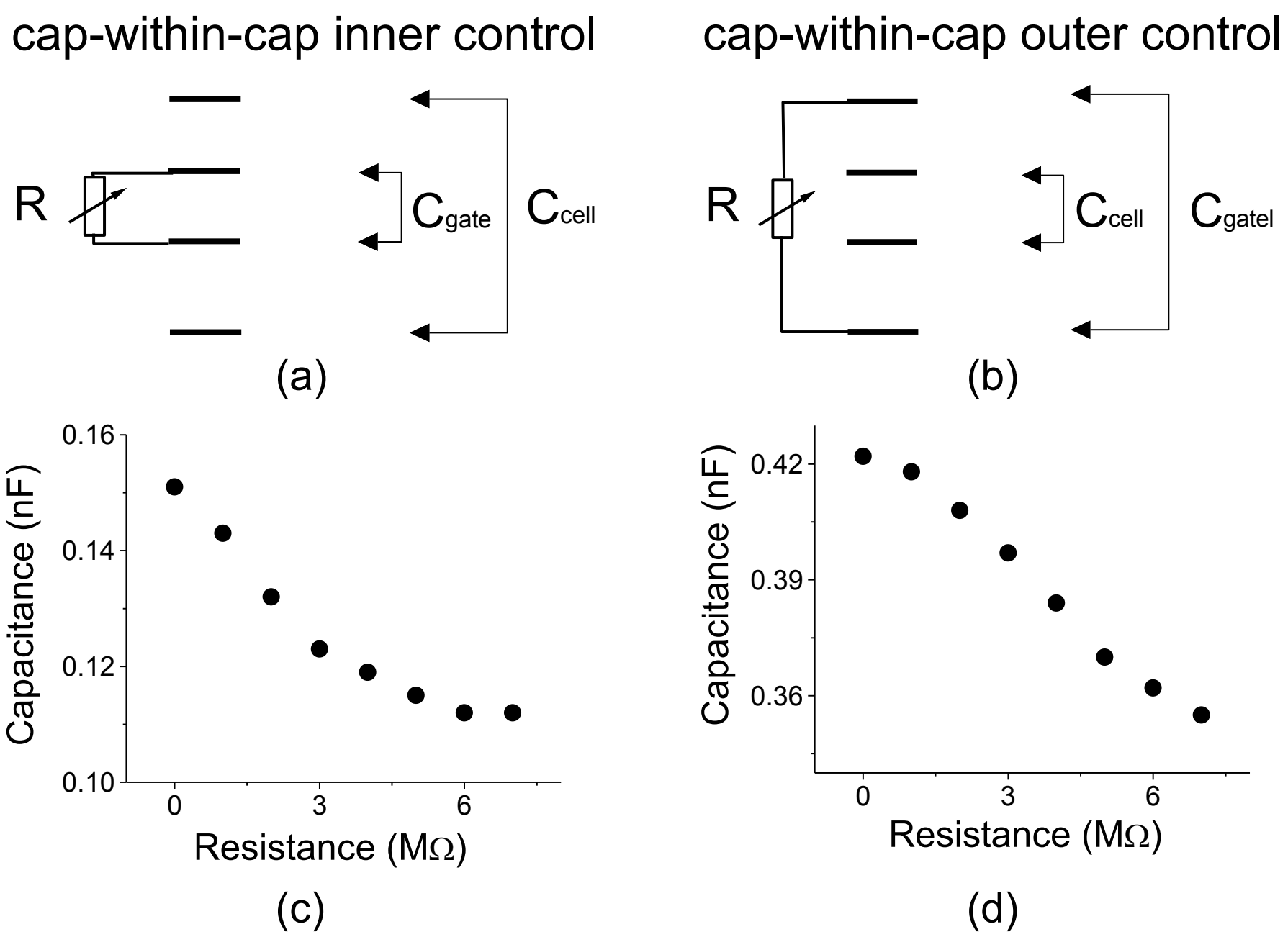

Fig. 5. The experimental configuration for an inner control (a) and an outer control (b). (c) Cell's capacitance as a function of resistance; inner gate control. (d) Cell's capacitance as a function of resistance; outer gate control. C=0.29 nF when the gate cap was open.

The scalability of the design is demonstrated by replacing the 'ordinary' plastic dielectric layer with ionic liquid. A 90 micron thick lens tissue was soaked with the ionic liquid for 30 minutes. The results are summarized in Table 1. Cell's capacitance has increased when the gate is short. Note the factor of ca 100,000 when the ionic liquid replaces the 'ordinary' dielectric.

| Gate Control | Gate Status | Cell's Capacitance (μF) |
|---|---|---|
| inner | open | 23.6 |
| inner | short | 34.5 |
| outer | open | 66.4 |
| outer | short | 93.2 |

Table 1: Electrolytic CWC with ionic liquid

***III.c. AC measurements:***

The overall capacitance of the cell may be assessed through the rise, or decay time of a 10 KHz square wave. The signal was applied to the outer terminals (Figs. 6a, 7a1-a3) using a function-generator. A 1 KΩ resistor was connected in series between the function-generator and the positive outer terminal leads. A reference experiment is shown in Fig. 6b. For the gated experiments, the same square wave was used as a gating signal, and was applied to the inner capacitor. Results are shown in Fig. 7b. If the polarity of the gate counters the polarity of the cell capacitor (Fig. 7a2), then, the time constant and hence the overall cell's capacitance increases (Fig. 7b2). If the polarity of the gate capacitor is in same the direction as the cell's polarity, then, the overall cell's capacitance decreases (Fig. 7b3).

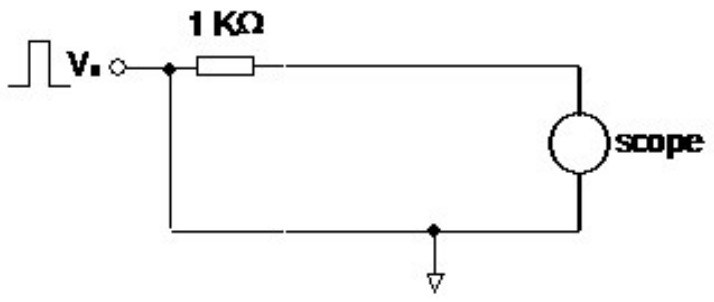

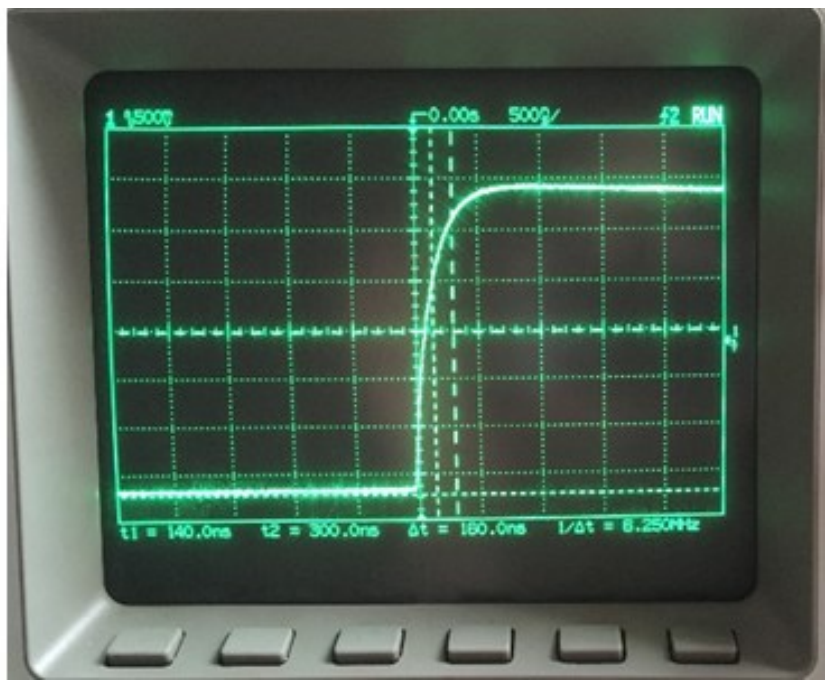

(a) (b)

Fig. 6. Reference data. (a) The circuit. (b) Rise-time of a square wave without the structured capacitor indicates some system capacitance. Dielectric spacers were used.

Open and biased cases are shown in Fig. 8. They demonstrate how one may increase (or decrease) the overall cell's capacitance by re-structuring the CWC.

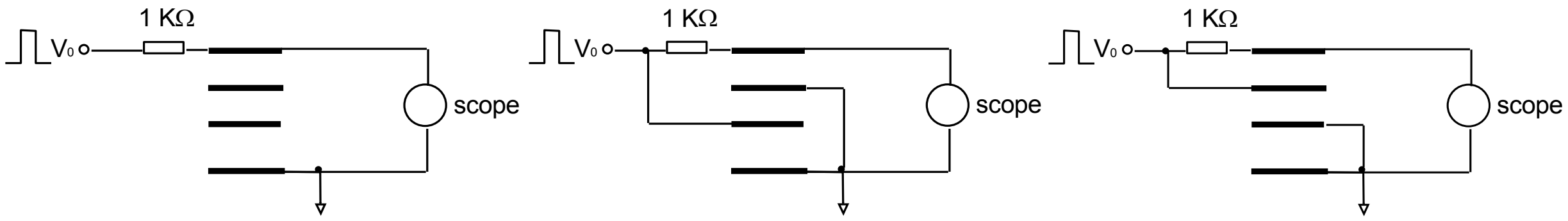

(a1) (a2) (a3)

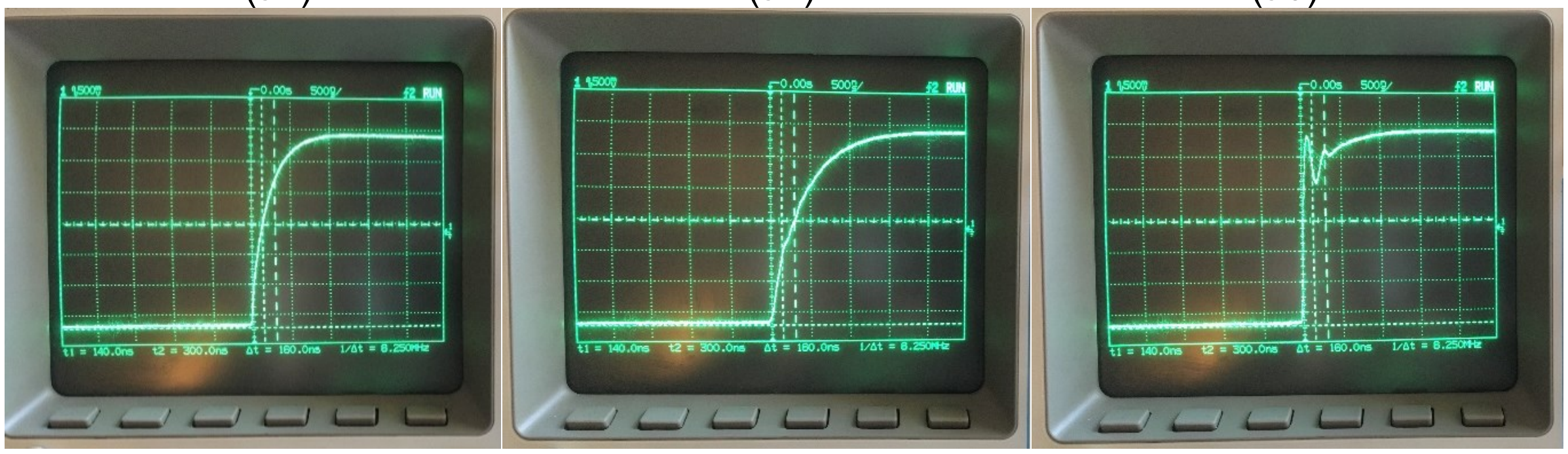

(b1) (b2) (b3)

Fig 7. (a1)-(a3) The circuits. (b1-b3) Traces of Voltage vs Time. (b1) The (inner) gate capacitor was open; (b2) the polarity of the (inner) gate capacitor was opposite to the polarity of the (outer) cell capacitor and the rise time-constant increased; (b3) the polarity of the (inner) gate capacitor was in the direction of the cell's polarity and the rise time-constant decreased.

## IV. Discussion:

If we consider a circuit approach to an inner construction with floating electrodes (Fig. 8), then the overall capacitance is made of three capacitors in series. Assuming that all capacitors are of similar value, then, C+C+C=C/3. This is similar to an ordinary parallel plate capacitor whose spacing is 3 times larger than for a single capacitor C. Thus it seems that we do not have any advantage by complicating the simple structure. Yet, along with Fig. 8b, if the inner gate is short, then the overall capacitance is C/2, or 50% larger than either the ordinary or the open and structured configuration. Outer gate construction (Fig. 8c) also results in a 50% capacitance increase; when the gate is short, we have one capacitor in parallel to two capacitors in series.

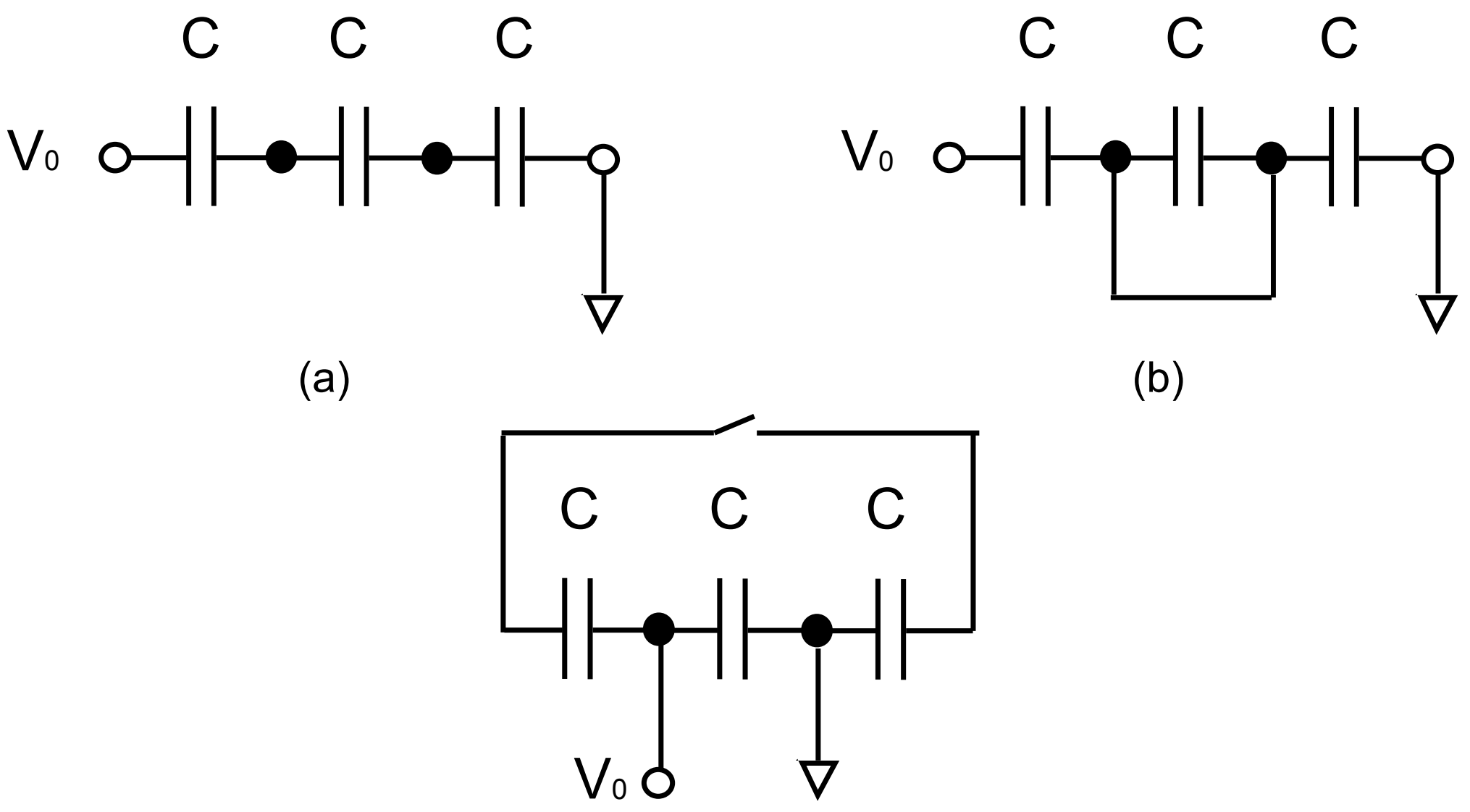

Fig. 8. (a) The equivalent circuit for inner gate construction results in a total capacitance of C/3, similar to an ordinary parallel plate capacitor. (b) Short inner gate with a floating potential diagram exhibits a total capacitance of C/2 compared with C/3 of the ordinary or 50% capacitance increase. (c) Outer gate construction result in a capacitance increase of 50%, as well.

In electric field terms, the increase in the cell's capacitance is attributed to the polarity of the gate structure, either for inner or outer construction. The gate dictates the polarity of the electrolyte filling the cell (the dielectric constant, so to speak). The larger this polarity is, the larger would be the cell's capacitance. Such induced capacitance is similar to artificial dielectrics [2-4] – dielectrics that are filled with floating metallic or semi-conductive objects to affect their permittivity values. Put it differently; the gate capacitor draws counter charges to its electrodes resulting in more electronic charges drawn to the cell's electrodes.

Unlike what may be deduced from equivalent circuitry, field simulations indicate that the particularities of the gate permittivity only marginally affect the overall cell's capacitance. For example, varying the permittivity value for an inner gate structure by three orders of magnitudes, from 1 to $10^3$ changes the overall cell's capacitance by less than 1%. That means that the larger effect is related to the induced surface charges across the gate electrodes and not to its capacitance.

The advantage of such multiple-electrode structure is clearer when placing the cell capacitor in an outer gate. There, one take full advantage of the smallest spacing possible for the cell capacitor and increase it further by encapsulating it with an outer gate structure (Table 1). The equivalent circuit is again made by two capacitors in series connected in parallel to the third and therefore, the capacitance increase is 50% compared to a single element.

***IV.a Energy Considerations:***

The broader question is: what do we gain by complicating the simple parallel capacitor's design? This is a linear system after all; in the absence of heating loss, all energies invested in charge separation are added up. In general, and if the electrodes are kept at a given potential, $V_i$, the energy stored in an array of conductors is written as [1],

$$U=\tfrac{1}{2}\sum_{ij}C_{ij}V_iV_j=\tfrac{1}{2}\sum_i Q_iV_i$$

Here, $C_{ii}$, $C_{ij}$, $V_i$ and $Q_i$ are the coefficient of self-capacitance, the coefficient of inductance between conductors i, j, the potential on the $i^{th}$ conductor and the charge on conductor i, respectively. In our case, $U=½(Q_{cell}V_0+Q_gV_g)$ with $Q_{cell}$ and $V_0$, respectively, the charge and potential on the cell's electrode. These two quantities are not arbitrarily determined and there is a linear relationship between them.

Let us consider the case where the two gate electrodes are connected together by a wire (namely, are short). Then, the total charge on the gate is zero. Yet, the polarization by the cell displaces the gate charge and creates a charge separation across the gate. Let us denote the induced charge on the gate electrode as q. The charge on a stand-alone cell's electrode, $Q_0$, will be modified as a result of this induced charge: $Q_{cell}=Q_0+q$. The total stored energy has therefore increased by $qV_0$. Likewise, the cell's capacitance has increased: $C_{cell}=C_0+q/V_{cell}$. Note that $V_{cell}$ is usually kept constant. The energy increase is supplied by the external source at potential $V_0$ (an ideal voltage source is defined as a source that can provide any required current or, charge for that matter, at a given potential bias).

## V. Conclusions:

Adding gate electrodes to a cell capacitor may be designed so to increase the overall cell capacitance. This may be beneficial when the cell capacitor has a bias limit on its potential, e.g., (super)capacitors. Interfacing this structure with nonlinear elements (such as diodes, transistors, or even light activated materials) is an exciting possibility.

## References:

1. L.D. Landau & E.M. Lifshitz, "Electrodynamics of Continuous Media", (Volume 8 of A Course of Theoretical Physics), Pergamon Press 1960
2. R. Collin, “Field Theory of Guided Waves”, Wiley-IEEE Press, ISBN: 978-0-87942-237-0.
3. H. Grebel and P. Chen, Artificial dielectric polymeric waveguides: semiconductor-embedded films, Opt. Lett. 15 (1990) 667-669.
4. H. Grebel and P. Chen, Artificial dielectric polymeric waveguides: metallic embedded films. H. Grebel and P. Chen J. Opt. Soc. Am. A 8(4) (1991) 615-618

## Capacitor-within-Capacitor

H. Grebel

Electronic Imaging Center, NJIT, University Heights, Newark, NJ 07102. grebel@njit.edu

Supplemental Information

**Control over CWC by gate voltage: inner gate control.** Examples of the potential distribution across an ordinary and structured dielectric capacitors are shown in Fig. S1a. An ordinary capacitor exhibits a linear electric potential distribution across its structure (Fig. S1b) and constant capacitance as a function of $V_{gate}$ (Fig. 2c). If one includes an inner gate capacitor (Fig. S1c), then the electrical potential distribution changes and the overall cell's capacitance varies as a function of the gate potential. The overall cell's capacitance has increased by a factor of ca 3. The symmetric case, shown in Fig. S1c is superior to an asymmetric case where the gate of same size may be located off-center or occupy fraction of the cell's width.

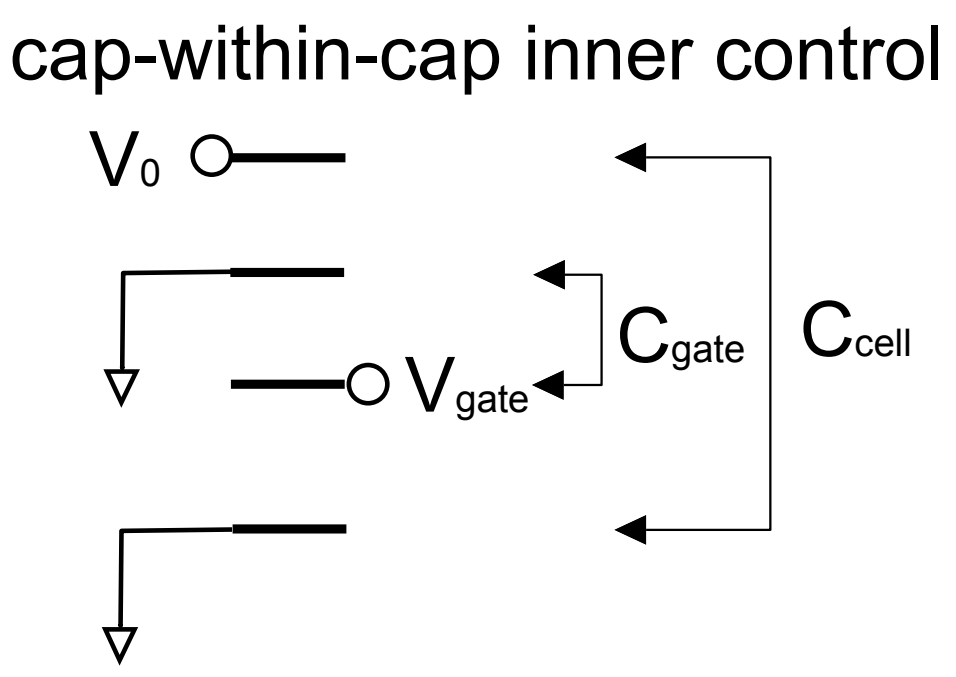

(a)

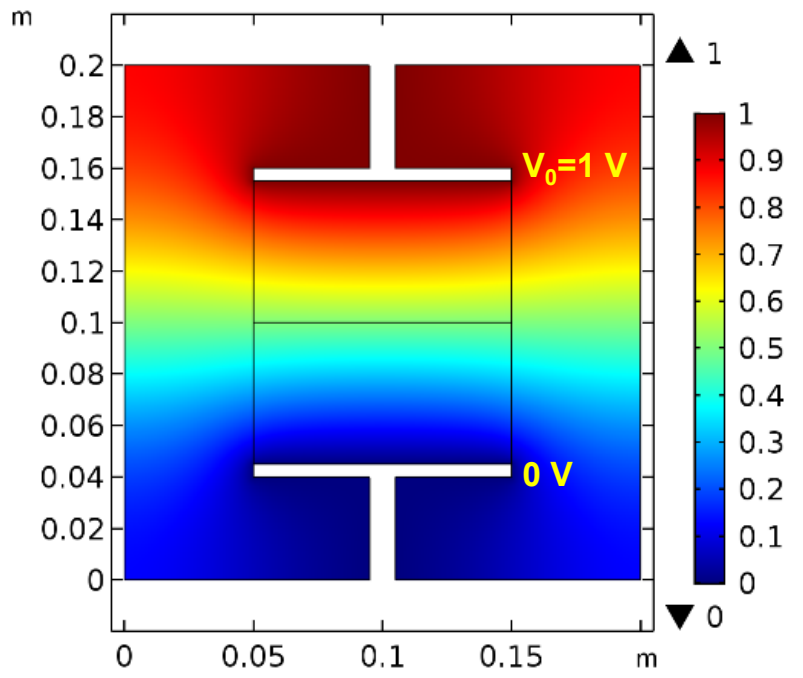

(b)

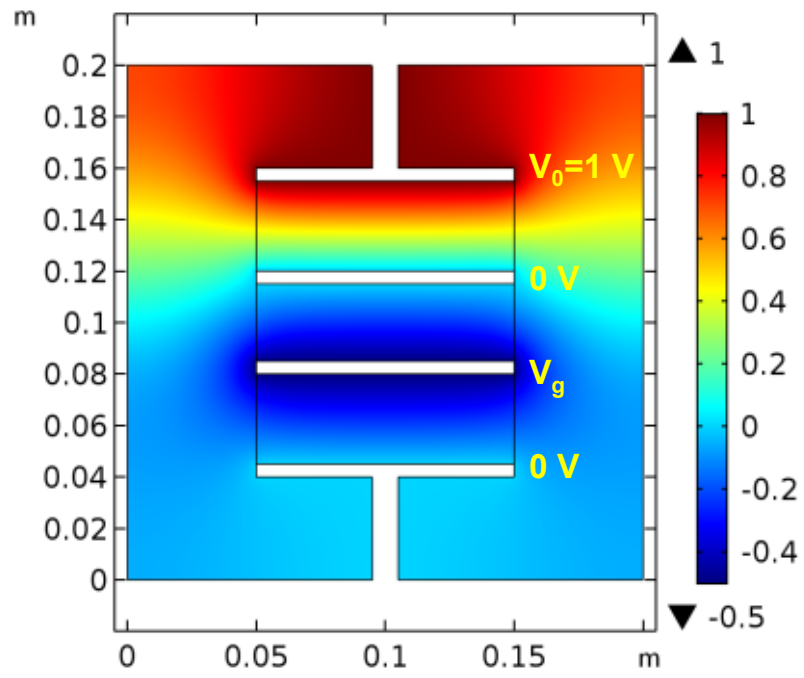

(c)

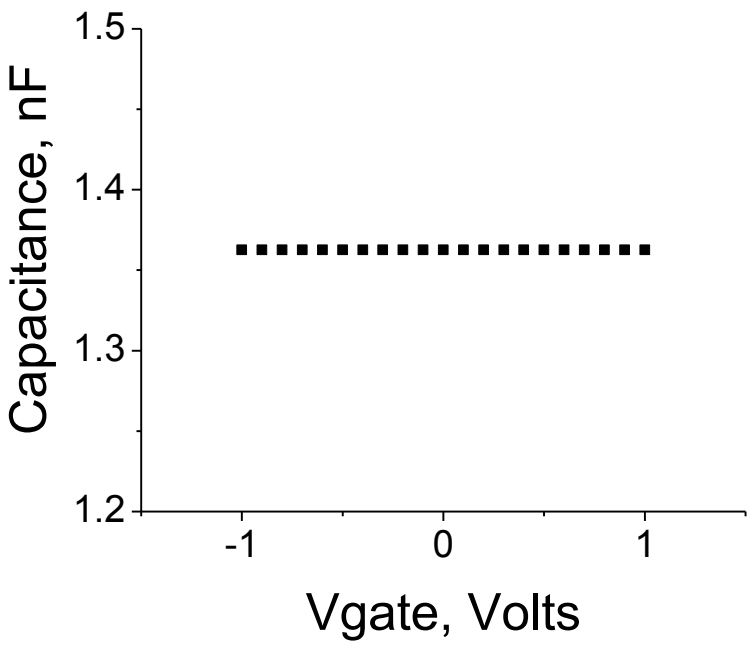

(d)

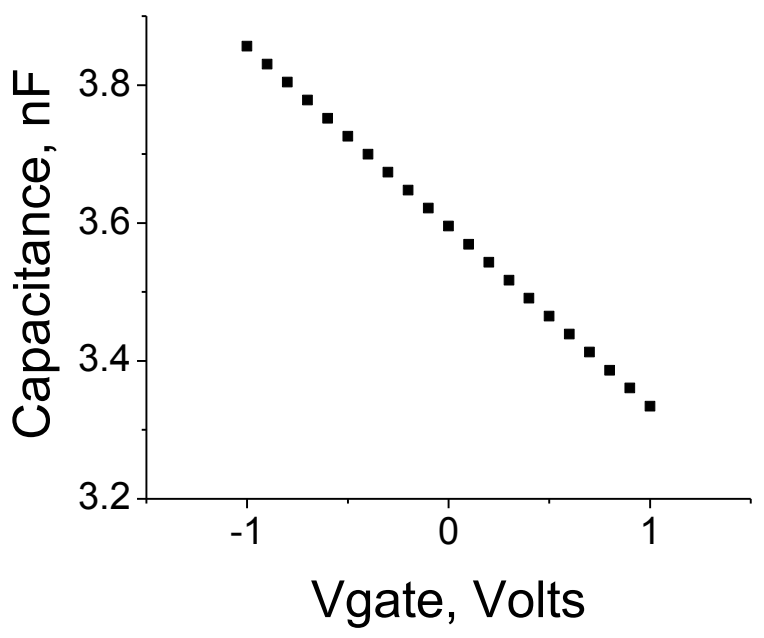

(e)

Fig. S1. (a) Ordinary capacitor: electric potential distribution when $V_0$=1 V. (b) Electric potential distribution when an inner capacitor, the gate capacitor is added. Shown is the potential distribution for $V_{gate}$=-0.5 V. (c) Without the inner gate capacitor, the capacitance (charge per unit cell's voltage) as a function of the gate voltage is constant (there is no gate). (d) The capacitance is linearly varying as a function of $V_{gate}$.

**Control over CWC by gate voltage: outer gate control.** Here the roles of the cell and the gate are interchanged. The dielectric capacitor is shown in Fig. S2. The inner cell capacitor ($C_{cell}$) is polarized when a potential difference $V_0$ is biasing the cell's electrodes. The controlling outer capacitor (the gate capacitor) is polarized by biasing it with $V_{gate}$. The structure was assumed to be immersed in a liquid dielectric with a relative dielectric constant of $\varepsilon_r$=100. The overall cell capacitance here is larger than the one depicted in Fig. S1e due to smaller cell's spacing; nevertheless, the cell capacitance is controllable by $V_g$.

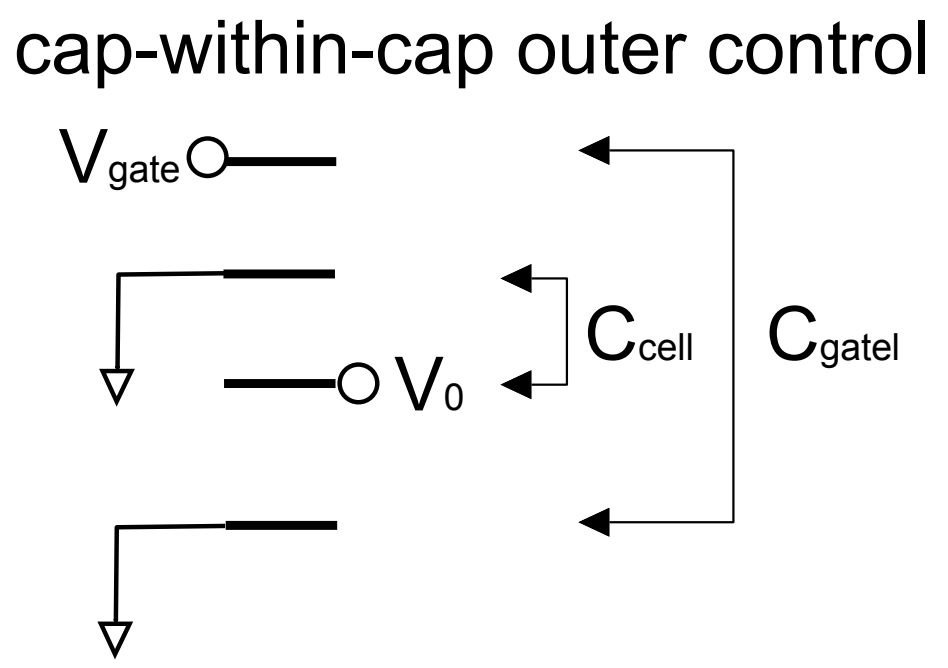

(a)

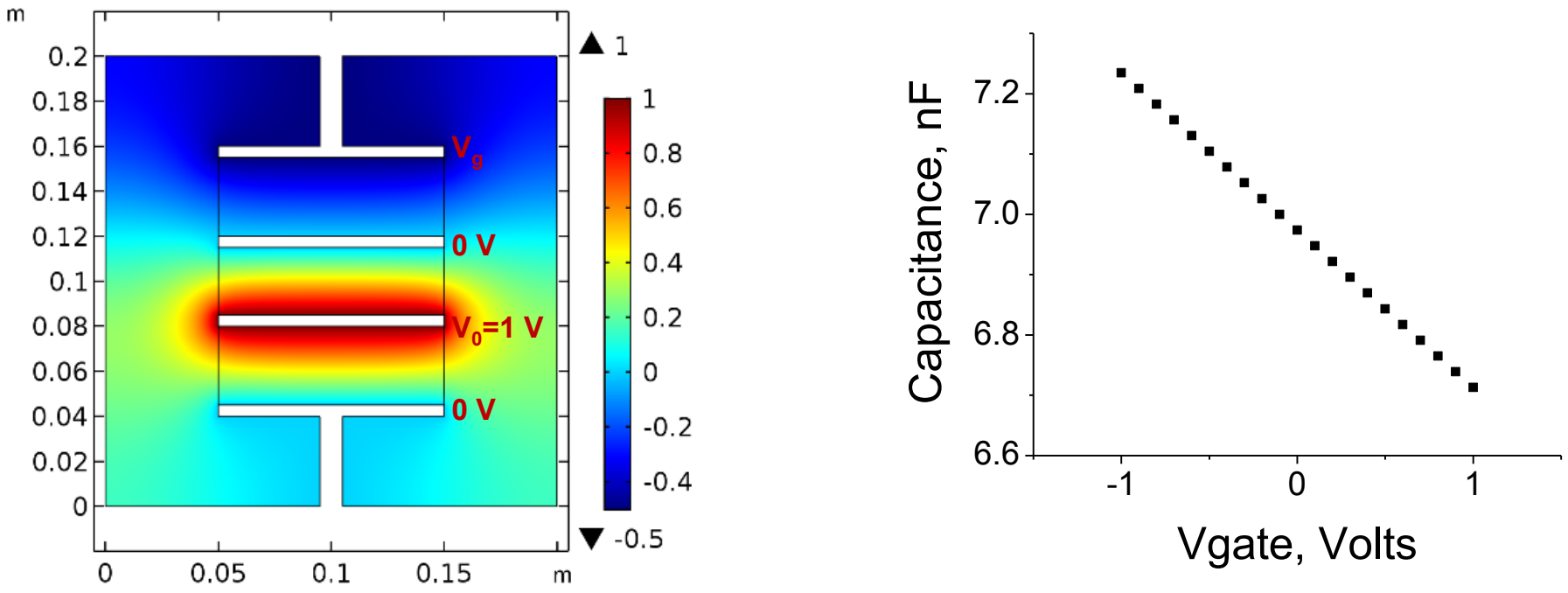

(b) (c)

Fig. S2. (a) The cell capacitor is enclosed by an outer gate capacitor. (b) Electric potential distribution when the outer capacitor serves as a gate. Shown is the potential distribution for $V_{gate}$=-0.5 V. (c) The capacitance is linearly varying as a function of $V_{gate}$, yet its change is smaller than the inner configuration of Fig. 2d.

**(Super)capacitors:** A (super)capacitor is an electrolytic capacitor, which has large electrode surface area [1-3]. A (super)capcitor takes advantage of the high capacitance at the interface between its electrodes and the electrolyte, across the Helmholtz's double-layer. In an ordinary (super)capcitor, the ionic charges are accumulated near the electrode surface and screen the electronic charges on the biased cell's electrodes. Their internal electric field counters the external field from the biased cell's electrodes. The total electric field in mid-region is almost zero and hence, with little, or no potential change. A highly conductive slab under an external field exhibits similar dipolar behavior: a large potential drop occurs at the slab ends with little potential drop in the mid-region. In the following, (super)capacitor was modeled by a capacitor, filled with a conductor (Fig. S3a,b). The potential of the conductor was floating. The ends of the conductor region were placed at close proximity to the electrodes of the cell. If the cell's electrodes are biased, then each electrode will draw counter charges to the end of the conductor, thus forming two large capacitors between electrode and conductor. This model is a simplification of a polarizable conductive model [4], used for a different application. The outer gate was simulated as a dielectric capacitor which encapsulates the (super)capacitor (Fig. S3c). Unlike a dielectric capacitor, the size of the (super)capacitor should not impact its overall capacitance; the large capacitance is mostly controlled by the width of the Helmholtz double-layer. One can easily verify our model by decreasing the overall capacitor cross section (Fig. S3b vs Fig. S3a). There was no capacitance change for these two cases and as can be seen from Figs. S3, the mid (super)capacitor region was maintained at a constant potential (no potential change).

A typical double-layer width is of nanometer scale whereas the separation between the anode and cathode of the cell or between the gate electrodes is larger; on a micrometer scale. That size discrepancy affects the computational complexity. Since the capacitance is directly proportional to the dielectric constant and inversely proportional to the electrode separation one may compensate for the small double-layer width by filling it with a material whose effective dielectric constant is very large (say, $\varepsilon_r=10^{10}$). As argued earlier, the choice of the gate dielectric is less important than the charge accumulated on the gate electrodes. For example, in Fig. S2, the relatively large spacing between the gate electrodes has been simulated through a small effective dielectric constant ($\varepsilon_r=10^5$). Similar cell capacitance were obtained with a relative gate permittivity value of $\varepsilon_r=10^8$. A substantial capacitance increase is only obtained when the dielectrics for the gate and cell reach comparable values.

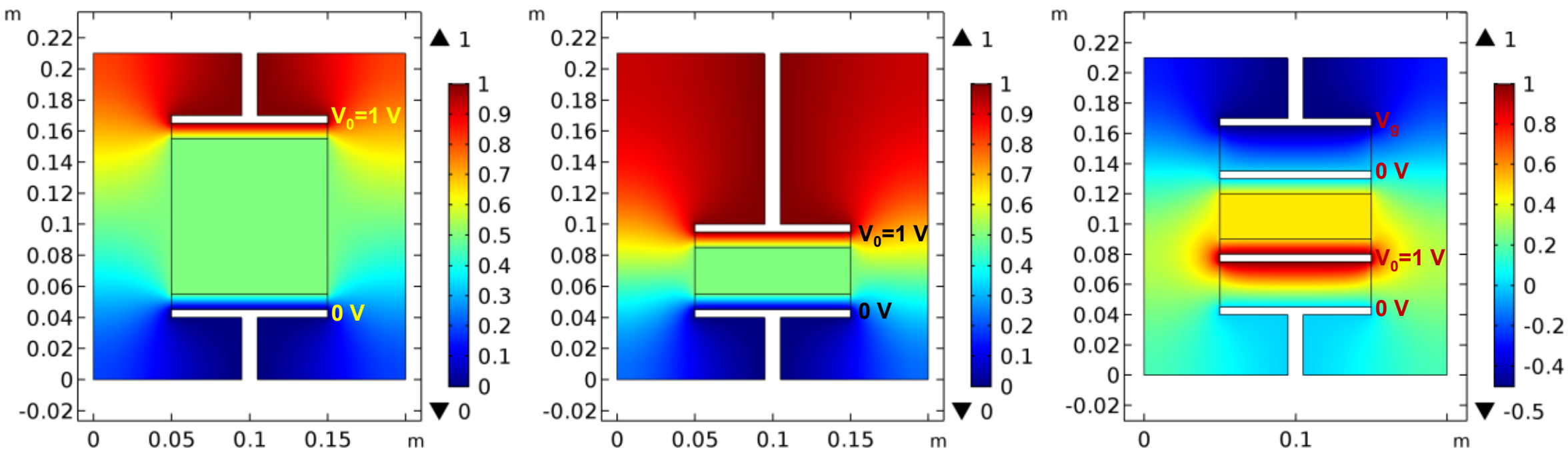

(a) (b) (c)

Fig. S3. (a) An ordinary (super)capacitor was modelled as a dielectric capacitor filled with a conductor at floating potential. Two large capacitors are formed at the ends of the conductor region. The potential is uniformly distributed within the conductor (mid-region of the (super)capacitor). (b) The (super)capacitor's cross section shouldn't affect it capacitance since the large capacitance is mostly limited to electrode/electrolyte interface. (c) Outer gate construction. The cell's capacitor is defined between the inner 0 V and $V_0$=1 V electrodes. The ordinary outer gate capacitor encapsulates the cell. The electrical potential distribution is shown for a gate bias of $V_g$=-0.5 V. Ground is represented by 0 V. The structure was immersed in an electrolyte. The potential distribution outside the cell is the result of fringe effects.

The control of the (super)capacitor by the outer gate voltage is shown in Fig. S4. There is a capacitance enhancement of ca 7.5% at $V_g$=0 V when compared to an ordinary (super)capacitor. Experimental data confirmed the increase in cell capacitance for structured design [5].

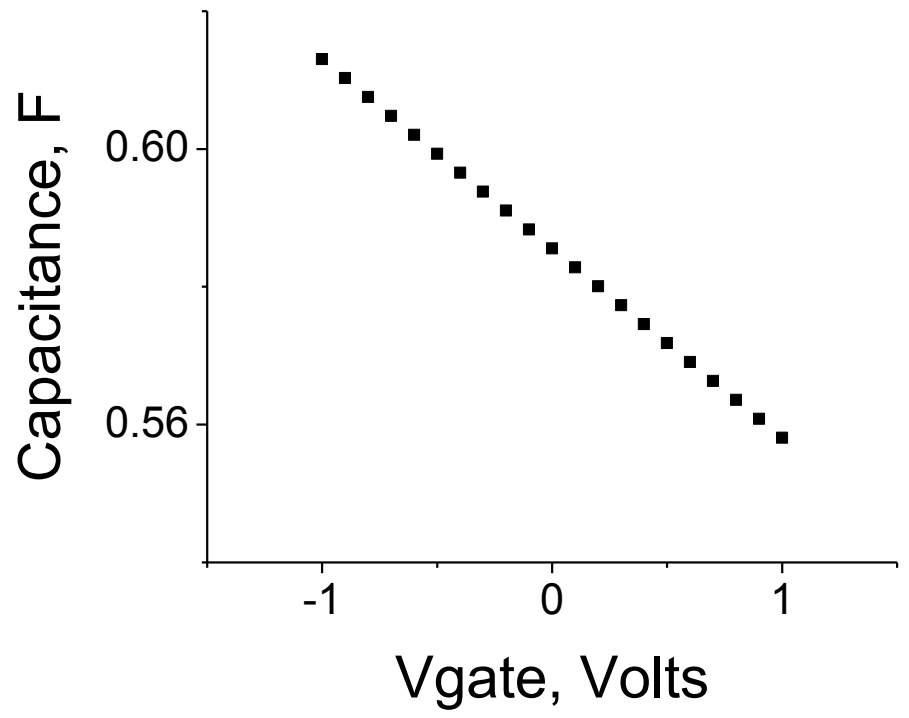

Fig. S4. Cell capacitance as a function of gate bias for Fig. S1c. The effective dielectric constants of the double-layer and the gate were $\varepsilon_r=10^{10}$ and $\varepsilon_r=10^5$, respectively. The overall capacitance at $V_g$=0 V was 7.5% larger than the capacitance of an ordinary (super)capacitor (Fig. S3b). More specifically, the values for the structured and ordinary capacitors were 0.58 F and 0.54 F, respectively.

**Applications:**

A typical (super)capacitor is fabricated by depositing an anode and a cathode directly on an electrolyte-soaked separator membrane. This structure is rather compact. Under these circumstances an outer gate construction may be more appropriate. At the same time, we proposed in the past to replace the otherwise insulating separator layer with a conductive double layer, which, according the prior discussions ought to increase the overall cell capacitance. In general, electrolytic (super)capacitors do not handle large potentials well. The proposed design may store more energy at a given low voltage. Controlling an electrochemical cell by intermediate gate structure have been proposed in the past [6-8].

Another application is the introduction of a diode (Fig. S5) either in series with $V_{in}$ or in series with $V_{out}$ (as shown) to affect the rise or fall time of the output signal. $V_{in}$ is a 1 $V_{p\text{-}p}$, square wave at frequency f=10 Hz. The dielectric is made of ionic liquid.

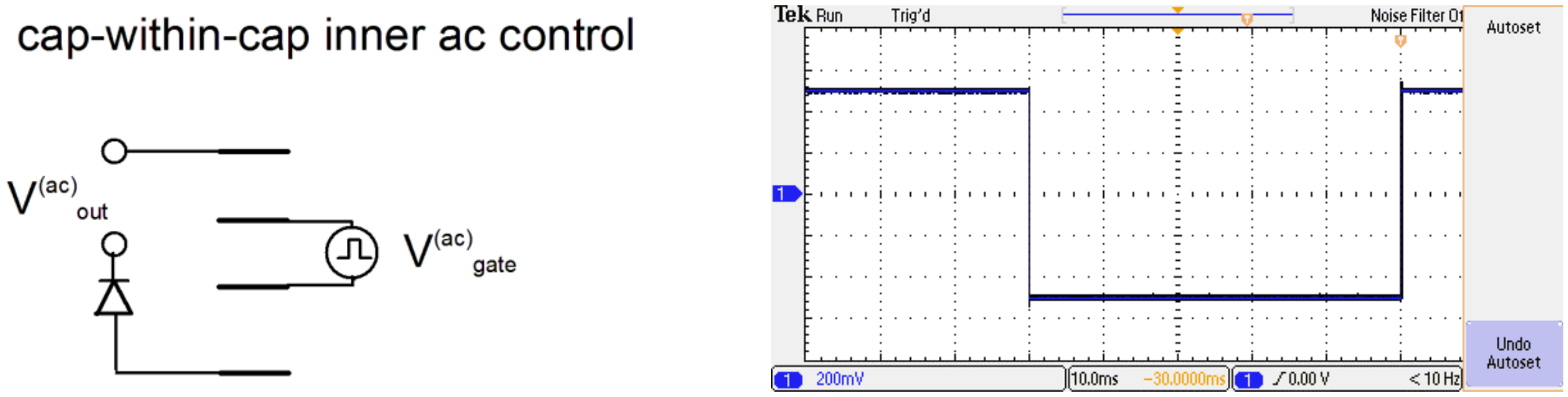

(a) (b)

(a) Schematics of the experiment. (b) The input signal

The direction of the diode dictates whether the rise or the decay time will be longer.

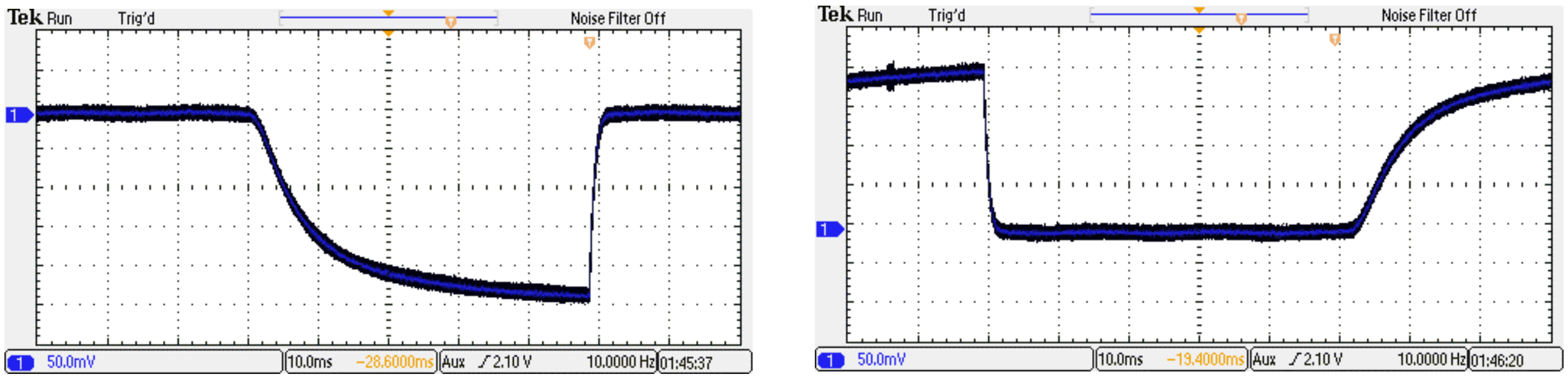

(a) (b)

$V_{out}$ in series with a diode: (a) Forward connection. (b) Backward connection

1. John R. Miller and Patrice Simon, Electrochemical Capacitors for Energy Management, Science, 321 (2008), 651-652.

2. Zhong-Shuai Wu, Guangmin Zhou, Li-Chang Yin, Wencai Ren, Feng Li and Hui-Ming Cheng, Graphene/metal oxide composite electrode materials for energy storage, Nano Energy, 1 (2012) 107-131.

3. R. Kötz, and M. Carlen, Principles and applications of electrochemical capacitors – ScienceDirect, Electrochimica Acta, 45 (2000) 2483.

4. Yu Takano, and K. N. Houk, Benchmarking the Conductor-like Polarizable Continuum Model (CPCM) for Aqueous Solvation Free Energies of Neutral and Ionic Organic Molecules, J. Chem. Theory Comput., 2005, 1 (1), pp 70–77. DOI: 10.1021/ct049977a

5. T. Selim Chowdhury and H. Grebel, Electrically tuned super-capacitors, https://arxiv.org/pdf/1512.08000. Recently, we have shown that adding a p-n layer junction to the, otherwise bare and insulating separator membrane, increases the cell capacitance by at least 10%
6. Haim Grebel and Yan Zhang, Controlling Ionic Currents with Transistor-like Structures, ECS Trans. 2 (2007), 1-18; doi: 10.1149/1.2408981
7. Amrita Banerjee and Haim Grebel, On the stopping potential of ionic currents, Electrochem. Commun. (2010), doi:10.1016/ j.elecom.2009.12.013.
8. Joel Grebel, Amrita Banerjee and Haim Grebel, Towards bi-carrier ion-transistors: DC and optically induced effects in electrically controlled electrochemical cells, Electrochimica Acta, 95 (2013) 308–312.